\documentclass{llncs}

\ifx\pdfoutput\undefined
\usepackage{graphicx}
\usepackage[dvips]{epsfig}
\else
\usepackage[pdftex]{graphicx}
\fi

\begin{document}
\frontmatter          % for the preliminaries
\pagestyle{headings}  % switches on printing of running heads

\mainmatter

\title{Service-oriented Communities:\\
Visions and Contributions\\
towards Social Organizations}

\author{Vincenzo De Florio and Chris Blondia}
\institute{University of Antwerp\\
        Department of Mathematics and Computer Science\\
        Performance Analysis of Telecommunication Systems group\\
        1 Middelheimlaan, 2020 Antwerp, Belgium. \vspace*{3pt} \\
%\and
        Interdisciplinary Institute for Broadband Technology\\
        8 Gaston Crommenlaan, 9050 Ghent-Ledeberg, Belgium.}

\maketitle

\begin{abstract}
%\centerline{\includegraphics[width=0.84\textwidth]{fname.png}}

With the increase of the populations, resources are becoming scarcer, and a smarter way to make use of
them becomes a vital necessity of our societies. On the other hand, resource management is traditionally
carried out through well established organizations, policies, and regulations that are often considered
as impossible to restructure. Our position is that merely expanding the traditional
approaches might not be enough. Systems must be radically rethought in order to
achieve a truly effective and rational use of the available resources. Classical concepts such as demand
and supply need to be rethought as well, as they operate artificial classifications that limit the true
potential of systems and organizations. 
Here we propose our vision to future, ``smarter'' systems able to overcome the limitations 
of the status quo. 
An example of such systems is the social organization that we call Service-oriented Community,
which we briefly describe.
We believe that such organizations---in heterarchical coexistence with traditional systems---provide 
the features necessary to prevent societal lock-ins like the ones we are experiencing in
assisting our elderly ones.
\end{abstract}

%%%%%%%%%%%%%%%%%%%%%%
\section{Introduction}
%%%%%%%%%%%%%%%%%%%%%%

With the increase of the human population, resources are becoming scarcer, and a 
smarter way to make use of them becomes a vital necessity if we want to get rid 
of or at least postpone unmanageability and chaotic behaviours. Assistance of 
the elderly population is a typical example: The share of the total population 
older than 65 is constantly increasing worldwide~\cite{Euro04,Gou03}, while 
the current organizations still provide assistance in a non-efficient, 
inflexible way. Though effective when the context was different
and a large amount of resources was available to treat a smaller demand,
this approach is now becoming too expensive and thus unacceptable.
Merely expanding the current organizations without properly restructuring them is simply not working anymore.

As remarked already in~\cite{DeBl08a},
another case of this ``syndrome'' can be found in other domains as well,
e.g. in network software engineering. Let us consider
the software principle
of layered design---dealing with the problem of an ever growing
design complexity by decomposing functionality into specialized layers.
This strategy proved very effective in the infancy of the Internet,
when hosts where limited in number and static in nature. 
The current scenario of a predominantly mobile Internet pervading all aspects of
human society including goods and environments brought about
new unprecedented requirements that are hardly compatible with
the current needs for flexibility, adaptability, and personalized behavior.
The extra performance granted by technology improvements is often
wasted or under-utilized if we do not restructure the software
architectures they are embedded into.

Business entities and even societies are no exception to this
trend. Indeed, often such systems were built under similar
``relaxed'' conditions---market and demographic contexts that were much less
stringent than the current ones. It is then no surprise that,
even though ever increasing amounts of resources are being pumped up into
these organizations, still they experience congestion and at times
fail to meet their expected quality of service.

A thorough analysis of the reasons behind these inefficient organizations
is beyond the scope of this text; what we would like to remark here is that
one of the factors that most likely play an important role here is that of the so-called
``lock-in'', defined by Stark as ``the process whereby early successes
can pave the path for further investments of new resources that eventually
lock in to suboptimal outcomes''~\cite{Stark}. 
When applied to society and
its organizational structures, lock-ins represent the loss of the ability
to rethink or at least revise well established services such as health-care.
Interestingly enough, Stark refers to this ability as \emph{adaptability}. A system
(be it e.g. a computer system, or a business entity, or a societal service)
is called by Stark adaptable when it is able to actuate ``ongoing reconfigurations of organizational
assets''. Lock-ins are the result of a loss of adaptability,
that is the loss of the ability to 
\emph{innovate\/} (evolve, best-fit etc.) \emph{through recombination\/}~\cite{Hol95}.

In what follows we propose our vision to future, ``smarter''
systems able to overcome the limitations of the status quo. Our position is that
such systems require what Boulding called
``gestalts''~\cite{Bou56}, namely concepts able to ``directing research towards the gaps 
which they reveal''. In this
paper we elaborate on this and show how such gestalts can pave the way towards novel reformulations of
traditional services able to reach a better and more sensible management of the available resources and
cope with their scarcity. A way to achieve this is also briefly sketched 
as a generalization of our concept of a ``mutual assistance community''~\cite{SDGB10b}, which we
called Service-oriented Community.

The structure of this paper is as follows: First in Sect.~\ref{s:mac}
we briefly recall the design of our first mutual assistance community. Section~\ref{s:onrefl}
discusses guidelines, conceptual tools, and hypotheses of such socio-technical systems.
Next, in Sect.~\ref{s:soc}, we briefly introduce our Service-oriented Community and
show how this may be considered as an example of what Boulding calls 
a social organization~\cite{Bou56}.
Finally in Sect.~\ref{s:end} we draw our conclusions.

%%%%%%%%%%%%%%%%%%%%%%%%%%%%%%%%%%%%%%%%%%%%%%%%%%%%%%%%%%%%%%%%%%%%%%%%%%%%%%%%%%%%%%%%%%%%%%%%%%%%%%%
\section{An Exercise in Innovation Through Recombination: The Mutual Assistance Community}\label{s:mac}
%%%%%%%%%%%%%%%%%%%%%%%%%%%%%%%%%%%%%%%%%%%%%%%%%%%%%%%%%%%%%%%%%%%%%%%%%%%%%%%%%%%%%%%%%%%%%%%%%%%%%%%

\begin{figure}[t]
%\centerline{\includegraphics[width=0.98\textwidth]{soamac.png}}
{\includegraphics[width=1.0\textwidth]{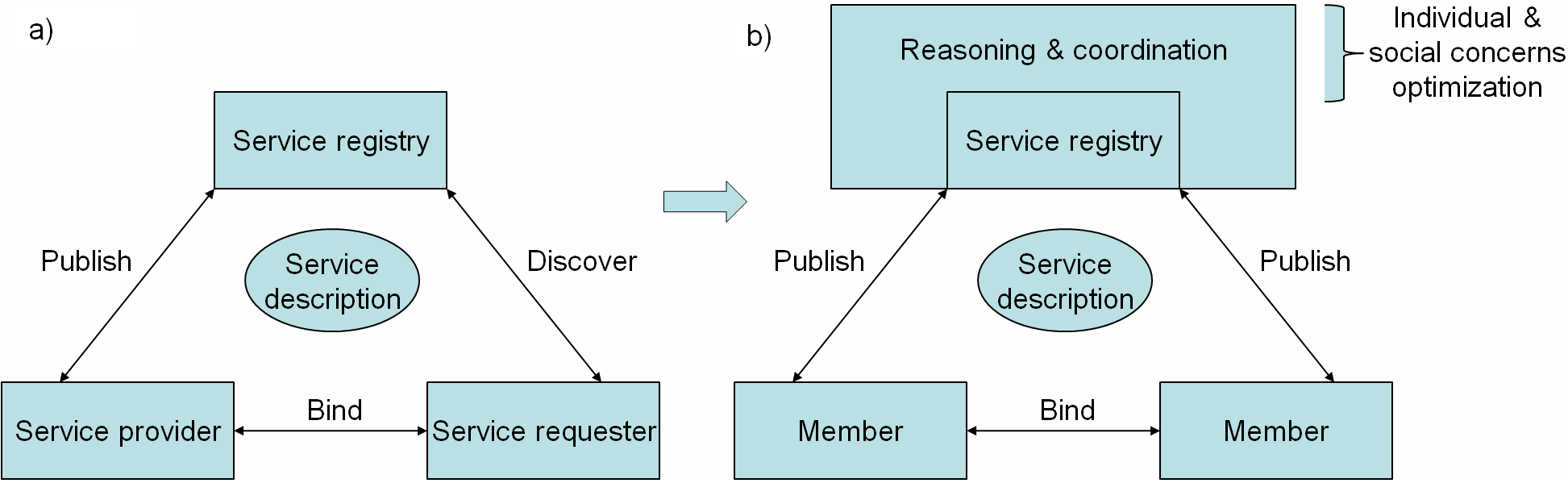}}
\caption{In a), the classical service-oriented model is recalled. In b), a simplified representation 
of the MAC model is displayed.}
	\label{f:soamac}
\end{figure}

What we call the mutual assistance community (MAC) is a service-oriented architecture~\cite{Erl05} coupling
services provided by smart devices with services supplied by human beings into an alternative
organization for AAL services. Semantically annotated services and requests for services are
published into a service registry (the coordination center) and trigger semantic discoveries of
optimal responses. Such responses are constructed making use of the
available resources and of the available context knowledge so as to optimize both individual and social
concerns. Figure~\ref{f:soamac} summarizes the peculiar differences between the classical service-oriented
model and that of our MAC.
A detailed description of the MAC may be found in~\cite{SDGB10b}.

The key idea behind MAC is that, in order to find an effective and also
cost-effective solution to big societal problems such as those addressed by AAL, the whole of society
must be included in the picture. When considering most of the available approaches to AAL (as surveyed e.g.
in~\cite{SDGB09a}), one may observe that a common aspect is often that people are divided into classes,
e.g. primary users (the elderly
people themselves), secondary users (carers) and tertiary users (society at large). This artificial
classification limits the effectiveness of optimally recombining the available assets into an effective
and timely response to requests for assistance. Furthermore, this classification into an active part of
society, able to contribute with worthy services, and a ``passive'' part only on the receiving side is
already a source of discomfort for people that are thus brought to feel they were \emph{once\/}
part of a society that now confines them to
a lesser state and dignity. Trying to reach emerging behaviours such as so-called e-Inclusion~\cite{eInc}
starting from
assumptions such as these is probably not the best course of action. When designing the MAC we
started from a different, more peer-to-peer approach in which people---be them elderly or otherwise---are
just members of a community---for instance the citizens of a small village or the people who subscribed
to a gym course. Members are \emph{diverse}, and this translates into a rich variety of services.
Diversity of course implies here different know-hows (e.g. those of a general practitioner, or of a
gardener, or of a retired professor of biology), different policies in providing their services (e.g.
well defined time schedule and associated costs, or dynamically varying availability to provide
free-of-charge services as occasional informal carers), different value systems, and so on. Another important characteristic here
is that members are not ``stationary'' but \emph{mobile}: They would ``wander around'' getting
dynamically closer to or farther from other members. When a request for assistance is issued, a response
can be orchestrated by considering the available members, their competence, their availability, and their
location with respect to the requester. A key aspect here is the ability to reason in an intelligent way
about the nature of the requests and that of the available assets. Unravelling analogies through semantic
reasoning~\cite{BPGR08} promotes both a higher level of resource utilization and a stronger degree of
e-Inclusion~\cite{SDGB07a}.

In a nutshell, our MAC is a socio-technical system in which AAL services (human or otherwise) can be
queried, located and dynamically orchestrated. In a MAC, elderly people would not
always be passive receivers of care and assistance, but occasionally they could play active roles. As an
example, if member $A$ feels lonely and wants to have a walk with someone, while member $B$ feels like
having some physical exercise with someone, then the MAC is able to capture the semantic similarities of
those requests and realize that $B$ could be the care giver of $A$ and vice-versa---through the
so-called ``participant'' model~\cite{SDGB07a} of our system.
Societal resources can then be spared, at the same time also preserving human dignity.
Simulation~\cite{SDFB06a} indicates
that systems such as this---able that is
to intelligently exploit the dynamically available resources---have the
potential to reduce significantly societal costs at the same time increasing efficiency,
manageability, and e-Inclusion.
Evidence of the widespread of such ideas can
be found also in the text of the third call of the European AAL Joint Programme~\cite{AAL-3}, whose main
focus is on ``solutions for advancement of older persons' independence and participation in the
\emph{self-serve society}''.

\begin{figure}[t]
\centerline{\includegraphics[width=0.75\textwidth]{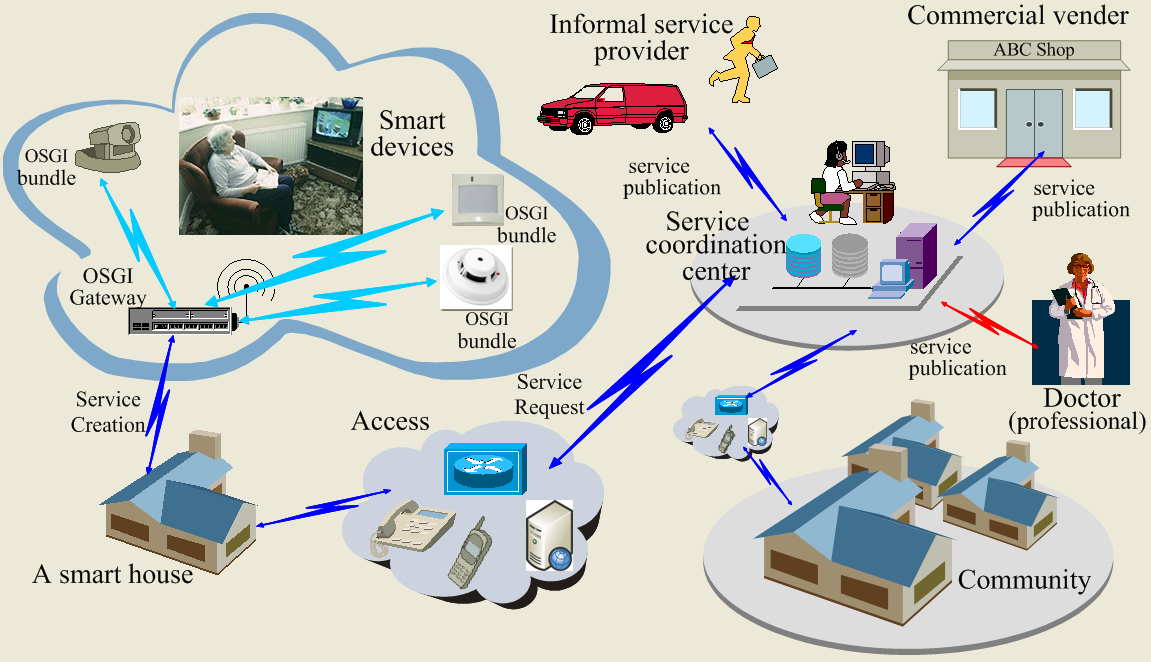}}
\caption{Representation of our prototypic OSGi-based implementation of a Mutual Assistance Community~\cite{SDGB10b}.}
	\label{f:mac}
\end{figure}

Figure~\ref{f:mac} sketches our prototypic implementation of the MAC based on the OSGi middleware.

%%%%%%%%%%%%%%%%%%%%%%%
\section{On Reflection}\label{s:onrefl}
%%%%%%%%%%%%%%%%%%%%%%%

In what follows we provide our reflections on two issues:  What guidelines and tools are
most needed when trying to devise adaptable organizations? And what are the necessary underlying
hypotheses of organizations such as our MAC?

%-------------------------------------------------%
\subsection{Conceptual Tools and Design Guidelines}
%-------------------------------------------------%

As already stated, 
we refer to our MAC as an exercise in the above mentioned ``innovation through recombination'': No new
assets are on purpose devised in our community; instead, a new organization of already existing assets is
proposed where the inherent potentials of our societies (mobility, diversity, density, etc.) are
exploited in a different way to the benefit of society itself.

What are the main lessons we can derive from our exercise? How can we generalize it and use it as a
model to tackle the problem of other societal lock-ins? We found an insightful guideline to answer these
questions in the work of Kenneth Boulding and especially in his classical paper~\cite{Bou56}. In such a
relatively short text we found a number of important concepts that reverberate throughout our overall
research experience. In particular, Boulding provides his interpretation of General Systems Theory as a
unitary conceptual framework in which ``to point out similarities in the theoretical constructions of
different disciplines, where these exist, and to develop theoretical models having applicability to at
least two different fields of study''. Being able to capture the similarities in the theoretical
construction of two or more disciplines allows the theorist (or the designer) to identify what Boulding
calls a ``gestalt'': Concepts that are devised in the framework of a specific discipline but can find a
direct application or a direct analogue in another discipline. Gestalts are greatly important: As
disciplines in general evolve at different paces, identifying gestalts in related disciplines permits to
direct research towards gaps or local minimums. In our experience, gestalts also allow \emph{lock-ins\/}
to be identified and provide practical guidelines to treat them.

As an example, the concept of \emph{feedback loop\/} is a gestalt that can be identified in as different a
discipline as cybernetics, biology, control theory, computer architectures, and social
science~\cite{PVR07}. Organizations that are based on or make use of feedback loops are able e.g. to
structure their function in accordance with a subset of the current endogenous and exogenous
conditions---that is, a subset of the current 
\emph{context}\footnote{Context is defined in~\cite{Dey01} as 
               ``any information that
                 characterizes a situation related to the interaction between
		 humans, applications and the surrounding
                 environment.''}.
Such subset represents a choice of context variables that are deemed as ``sensible enough'' to steer
optimally the function of the system. Another important aspect in these context-aware organizations is
the type and quality of the response they can provide.

It is again Boulding's cited paper that provides us with a detailed analysis of classes of context-aware
organizations. The bottom level of Boulding's classification is given by so-called ``frameworks'' and
``clockworks''. Such systems have a ``predetermined, necessary motion''~\cite{Bou56} that is they are
quasi context-agnostic. Next level is the one of ``thermostats,'' or systems that ``move to the
maintenance of any given equilibrium, \emph{within limits}''~\cite{Bou56}. Such systems focus on a very
limited set of context variables and ignore all the rest, exactly as a thermostat does with any variable
other than monitored temperature. The only response they can exhibit is also intended to affect that same
variable only. Continuing his classification, Boulding proposes other organizations (``cells'',
``plants'', and ``animals''), each of which is characterized by a more sophisticated degree of feedback
loops, with a larger amount and quality of sensory and actuation apparatuses. It is only with the level
of ``human beings'' and especially with that of ``\emph{social organizations}'' that systems are able to
introspect, analyze and locate their limiting factors---that is, their \textbf{lock-ins}---and to some
extent can learn how to reconfigure and reshape themselves so as to face a dynamically varying set of
environmental conditions. In other words, at this level Boulding introduces Stark's adaptability. Our
conjecture is that this is the feature we need to seed or steer in our organizations in order to let them
face unprecedented harsh conditions such as the ones our society are experiencing today. Thanks to the
generality of this gestalt, we have been able to apply successfully this concept to two different
contexts: The mutual assistance community described in Sect.~\ref{s:mac} and a software framework called
ACCADA~\cite{GuD10+}. In the former, the subset of context variables and actuation actions was extended
so as to include both technological and social aspects and services. The latter case is an attempt to
encode the reconfiguration capability into the feedback loop process itself so as to create truly
adaptable component-based software architectures.

%%%%%%%%%A sensible question here is of course \emph{how\/} to design adaptable organizations. 

%--------------------------------%
\subsection{Underlying Hypotheses}
%--------------------------------%

A key aspect in the effectiveness of our scheme is that members of our communities must be successfully
motivated to use the service and to offer their own services---our studies show that the more this
happens, the more the general welfare of the system increases~\cite{SDGB07a}. From this we derive two
main ``lessons'': First, that systems such as these must be designed so as to adaptively restructure
themselves according to each and every user. This adaptive personalization would allow to overcome
``usability barriers'' such as the so-called ``grey digital divide''~\cite{Mil03} and foster the
participation of the greater part of society. Secondly, we observe how another important key to a
successful spread of a heterarchical system such as our MAC calls for an overturn of social negative
values. Unfortunately, often a clash exists between this need and those social techniques and trends that
favor personal profit over the public interest~\cite{HeCho88,Cho02,MuHCho09}. This model's indiscriminate
use of negative values to manufacture consent is often used even in politics (stir social division, draw
fake enemies, fight diversity etc.) and results in the widespread loss, embezzlement, or misuse 
of a sort of ``Social Energy''---\emph{viz.}, the self-serve, self-organization, and
self-adaptability potentials of our societies.

%%%%%%%%%%%%%%%%%%%%%%%%%%%%%%%%%%%%%%%%%%%%%%%%%%%%%%%%%%%%%%%%%%%%%%%%%%
\section{Extending the Concept: Service-oriented Communities}\label{s:soc}
%%%%%%%%%%%%%%%%%%%%%%%%%%%%%%%%%%%%%%%%%%%%%%%%%%%%%%%%%%%%%%%%%%%%%%%%%%

As already remarked, effective adaptable organizations are characterized by an ``ongoing reconfiguration
of organizational assets''. When trying to design socio-technical systems such as our MAC an important
requirement is then the ability to reorganize dynamically a set of computer-based and human-based
services. Service orientation~\cite{All06} becomes \emph{the\/} privileged choice to designing complex
adaptive systems, its main characteristic being a set of well-defined, standard-regulated policies to
recombine loosely coupled atomic functionality. In recent years such functionality were extended so
as to include human-based services~\cite{BPEL4People,WSHT,STD08}.

It does not come as a surprise, then, how such \emph{gestalt\/} is being successfully applied to several
and seemingly unrelated domains. The organization presented in Sect.~\ref{s:mac} is indeed just another
example of service-oriented architecture, in this case applied to AAL. A similar example is given by
openAAL~\cite{openAALiance10}---an open source service-oriented middleware also
addressing ambient-assisted living.

Our question here is: Would it be possible to extend an organization such as our MAC to other
socio-technical domains? Would it make sense? What would be the societal returns in so doing?

To provide our answer to such questions let us begin by introducing a real-life case, given by the
Belgian enterprise Cambio (www.cambio.be). Cambio (and possibly other similar enterprises) provide simple
car renting services to their clients. Unlike other big players in car renting, Cambio allows cars to be
rented with shorter notice and for reduced durations and rates, thus fine-tuning the traditional concept
of renting a car. Obviously this concept could be further extended with e.g. intelligent car sharing
policies: Instead of just renting one car to one person, the system could reason about incoming requests
such as ``user $u$ needs to leave from source $S$ and reach destination $D$ within time $t$'' and try to
build up an optimal schedule that match several criteria. Meaningful explicit criteria could include
cost, number of hops, speed, reliability, and so on. Implicitly, such system would also have an effect on
important social matters, such as pollution from fuel combustion, traffic congestion, road dimensioning
requirements, intermediate consumption~\cite{ic}, et caetera. Note that no new resource would be
especially required to provide such services---which actually were already common at the beginning of last
century when people used to advertise for a traveling companion to share expenses~\cite{GBB}. A novel
organization based on service orientation, custom semantic reasoning, and proper human computer
interaction devices, could provide the necessary socio-technical foundation for such a service.

It is our conviction that an organization such as our MAC could be easily tailored towards such new
service context. Moreover we believe that several other classes of services could be supplied by tailored
MAC's. Should our conjecture prove true, then it would be possible to conceive a sort of multi-purpose
system where devices and human beings with different capabilities, competence, and information could be
optimally orchestrated to devise intelligent responses to situations ranging from a stroke to a
earthquake or a fire, or to needs such as connecting people together to share knowledge or
collaboratively achieve a common goal. Such ``service-oriented community''
%, symbolically represented in Fig.~\ref{f:soc},
would make use of a higher level feedback loop to reorganize itself and its services to
better match the current context---from both a personal and a general, societal perspective. Our ACCADA
middleware~\cite{GuD10+} could be used to set up such ``meta-adaptive'' loop.

%\begin{figure}[t]
%\centerline{\includegraphics[width=0.60\textwidth]{soc.png}}
%\caption{A possible implementation of a Service-oriented Community.}
	%\label{f:soc}
%\end{figure}

%%%%%%%%%%%%%%%%%%%%%%%%%%%%%%%%%%
\section{Conclusions}\label{s:end}
%%%%%%%%%%%%%%%%%%%%%%%%%%%%%%%%%%

Currently resource management is mostly
carried out through traditional, well established organizations, policies, and 
regulations that are often regarded as immutable to the point that any 
restructuring is considered as unthinkable or even dangerous. The current trend 
to deal with this problem is to complement the existing systems with other ``compatible'' approaches 
based e.g. on information and communication technology. One such approach is 
given by the use of smart devices and houses for elderly people. Our position 
in this paper is that merely \emph{complementing\/} the traditional approaches 
be not enough. We believe that systems and their ecosystems must be radically 
\emph{rethought\/} if we want them to achieve a truly effective and rational use of 
the available resources. Concepts such as demand and supply, as well as roles such as
producer and consumer, need to 
be thoroughly revised too, as their artificial classifications prevent
systems and organization to achieve their true potential.
The peer-to-peer ``participant'' model~\cite{SDGB07a}
of our model provides a concrete example of this,
removing when possible the unnecessary distinction between care-givers and patients.

In this paper we have proposed our 
vision to future, ``smarter'' systems able to overcome the limitations of the 
status quo. Such systems require what Boulding called \emph{gestalt}, namely 
concepts able to ``directing research towards the gaps which they reveal''. In 
this text we elaborated on this and showed how such gestalts can pave the way 
towards novel reformulations of traditional services that are able to exhibit a better and 
more sensible management of the available resources and to cope with their 
scarcity. Our vision of a Service-oriented Community was also introduced as a 
generalization of our concept of a Mutual Assistance Community.
We believe that such communities---in heterarchical 
coexistence with traditional systems---provide the necessary diversity and 
innovation orientation to prevent societal lock-ins such as the ones we are 
experiencing today in assisting our elderly ones. 

Our position
is that the traditional, hierarchical
model of governmental- or enterprise-driven and -regulated services could be replaced by
a more ``heterarchical'' view~\cite{Stark,Roc01} of concurrent providers based on different approaches
and possibly different values and missions. Such a new model
would allow our societies to function as
Boulding's social organizations, in which ``the unit [\ldots] is not perhaps the person but the
\emph{role}---that part of the person which is concerned with the organization or situation in question.
Social organizations might be defined as a set of roles tied together with channels 
of communication''~\cite{Bou56}.
A dynamic management of such roles---as foreseen in our vision of a
service-oriented community and promised by recent trends and emerging concepts
such as Web 3.0 and the Internet of Services---is possibly a necessary condition towards turning our
societies into effective social organizations.

Finally, we would like to highlight how preliminary experiments and common sense
suggest that another important prerequisite to
truly reaching this highly ambitious goal is learning how
to channel the potential of our ``Social Energy'' to the true benefit of society itself.

\subsection*{Acknowledgments}
We like to acknowledge how
several of the key ideas discussed in here are the result of
many discussions with M. Tiziana Bianco.
Also our gratitude goes to Hong Sun and Ning Gui, with whom we carried out several of the
research activities discussed in this paper~\cite{SDGB10b,SDGB09a,SDGB07a,SDFB06a,GuD10+,GSDB07},
as well as to our reviewers for their useful suggestions.

\bibliographystyle{splncs}
%\bibliography{thesis}

\end{document}